\documentclass[twoside,fleqn]{article}
\usepackage{espcrc2,texdraw,epsfig}
\usepackage{amssymb,amsfonts,float,pifont}
\usepackage{amsmath}
\newcommand{\beq}{\begin{eqnarray}}
\newcommand{\eeq}{\end{eqnarray}}
\newcommand{\be}{\begin{equation}}
\newcommand{\ee}{\end{equation}}
\title{Topology and Confinement in SU($N$) Gauge Theories
\thanks{talk presented by B. Lucini}}
\author{B. Lucini$^a$ and M. Teper$^a$
\vskip 5mm
$^a$ Theoretical Physics Department, University of Oxford, 1 Keble Road,
Oxford OX1 3NP, UK}

\begin{document}

\begin{abstract}
The large $N$ limit of SU($N$) gauge theories in 3+1 dimensions is
investigated on the lattice by extrapolating results obtained for
$2 \le N \le 5$. A numerical determination of the masses of the
lowest-lying glueball states and of the topological
susceptibility in the limit $N\to\infty$ is provided. Ratios of the tensions
of stable $k$-strings over the tension of the fundamental string are
investigated in various regimes and the results are compared with
expectations based on several scenarios -- in particular
MQCD and Casimir scaling. While not conclusive at zero temperature in
D=3+1, in the other cases investigated our data seem to favour the
latter.
\end{abstract}

\maketitle
\setcounter{footnote}{0}

\section{Introduction}
A lattice investigation of the large $N$ limit of SU($N$) gauge
theories~\cite{largeN} is interesting in many respects: it can improve our
understanding of SU(3) gauge theory as a base for full QCD and  allows
a direct comparison with predictions derived from theories beyond the
Standard Model, often rigorous only when $N\to\infty$. 

Existing lattice results suggest that SU(2) and SU(3) may be close to the
large $N$ limit. For this reason, we have investigated SU($N$) gauge theories
in 3+1 dimensions for $N=2,3,4,5$~\cite{glue01}. If as $N\to\infty$ a smooth
limit exists, corrections to this limit can be expressed as a power
series in $1/N^2$. Of course, a priori it is unknown whether observables
at small $N$ can be continued analytically to $N=\infty$ and how many
terms of the series are needed to relate quantities at finite $N$
to their $N=\infty$ values.
To answer those questions, we have studied the dependence on $N$ of the
fundamental string (Sect.~\ref{secstring}), of the mass of the $0^{++}$ and
$2^{++}$ glueballs (Sect.~\ref{secglueballs}) and of the topological
susceptibility (Sect.~\ref{sectopology})~\cite{glue01}. 

As soon as $N=4$ other strings appear in the spectrum that are
stable~\cite{mqcd}. These strings are denoted as $k$-strings,
$k=1,...,N/2$ being an integer
identifying the class of the representations of the gauge group
under which sources connected by a $k$-string transform. The values of the
ratios of the string tensions $\sigma_k/\sigma_f$ (the denominator being the
tension of the fundamental string) give insights on the mechanism of colour
confinement. A numerical investigation of those
ratios~\cite{usstring,usstring2} is reported in Sect.~\ref{seckstrings}.
\section{The string tension}
\label{secstring}
In our calculations we have used the Wilson action
\(
S = \beta \sum_{\mu > \nu,i} \left( 1 - \frac{1}{2N}\mbox{Tr} \left(
U_{\mu \nu}(i) + U_{\mu \nu}^{\dag}(i)\right) \right)\),
with $\beta = 2N/g^2$, $g$ being the coupling constant.

The string tension has been extracted by looking at the decay of correlation
functions of blocked Polyakov loops. To the lowest order in $1/L$, the string
tension $\sigma$ is related to the smallest mass governing this decay by
the formula
\begin{equation}
\label{fluxmass}
m = \sigma L - c_s \frac{\pi (D -2)}{6 L} \ , 
\end{equation}
where $L$ is the length of the string and $D$ is the number of dimensions.
If the infrared behaviour of the string can be described by an effective
string theory, the coefficient $c_s$ is universal~\cite{Luscherstring}
and depends only on the number of bosonic and fermionic modes propagating
along the string.
A reliable extraction of the string tension must keep into account the term
proportional to $1/L$. For this reason, we have determined $c_s$
for the gauge group SU(2) in $D=2+1$ and $D=3+1$~\cite{usstring2}.
Our fits give $c_s = 1.066 \pm 0.036$ and $c_s = 0.94 \pm 0.04$
respectively. Those values are compatible at the 2$\sigma$ level  with
an effective string theory of the bosonic type and are incompatible with
other simple string models. In the following we will assume $c_s = 1$, i.e.
that the long range fluctuations of the string are described by an effective
string theory and that the string is bosonic.


The string tension is used to set the scale of physical quantities at
different $N$. Equivalent quantities are identified by equivalent couplings.
In the original 't Hooft's idea, equivalent couplings in the large $N$ limit
correspond to fixed $\lambda = g^2 N$.
We have verified that the string tensions for different $N$ lie on an
universal curve as a function of $\lambda_I = g^2_I N$, $g^2_I$ being the
tadpole improved coupling~\cite{glue01}. As a guidance, equivalent $\beta$ at
different $N$ can be identified by using the relationship
$\beta(N)/\beta(N^{\prime}) = N^2/N^{\prime 2}$.
\section{The mass spectrum}
\label{secglueballs}
With the method described in~\cite{glue01}, we have determined the mass
of the $0^{++}$ and $2^{++}$ glueballs at various $N$. Results are
available also for the $0^{++\star}$, but they are preliminary.

\begin{figure}[tb]
\begin{center}
\epsfig{figure=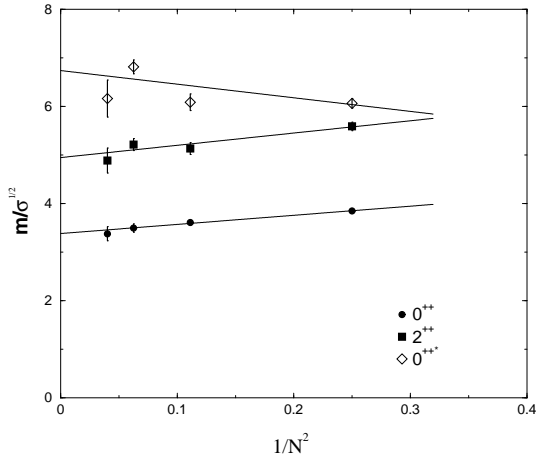, angle=270, width=7cm} 
\end{center}
\caption{The masses of the lowest-lying glueball states as a function
of $1/N^2$ with the best fits.} 
\label{fig:mass}
\end{figure}

Our analysis has firstly involved an extrapolation to the continuum limit for
each $N$, then an extrapolation to
the large $N$ limit of the continuum results. In Fig.~\ref{fig:mass}
we plot the masses in units of $\sqrt{\sigma}$ of the above states as a
function of $1/N^2$. A fit of the form $\hat{m}(N) = \hat{M} + a/N^2$ works for
all these states and all the way down to SU(2). Hence for these
quantities a smooth $N\to\infty$ limit exists and the first expected
correction explains the finite $N$ results all the way down to SU(2).
The $N=\infty$ values extracted for $\hat{M} = M/\sqrt{\sigma}$
are\footnote{In this work, our error bars refer to a confidence level of 68\%,
while in Ref.~\cite{glue01} a confidence level of 75\% was taken.} 
$\hat{M}=3.380(69)$ for the $0^{++}$, $\hat{M}=4.94(13)$ for the $2^{++}$
and $\hat{M}=6.74(16)$ for the $0^{++\star}$.

\begin{figure}[t]
\begin{center}
\epsfig{figure=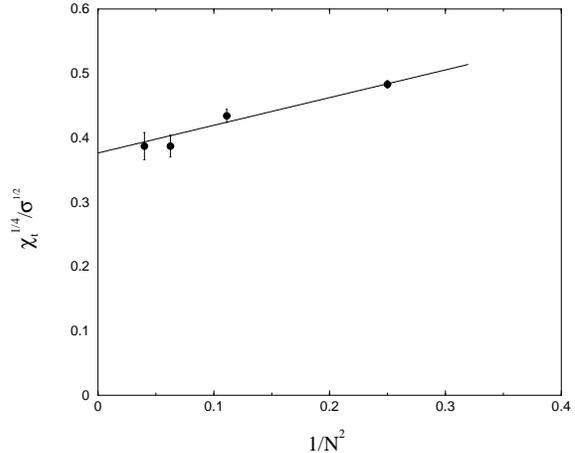, angle=270, width=7.4cm} 
\end{center}
\caption{The ratio ${\chi}_{t}^{1/4}/\sqrt{\sigma}$ as a function
of $1/N^2$ with the best fit.} 
\label{fig:chi}
\end{figure}

\section{Topological susceptibility}
\label{sectopology}
Large $N$ arguments were the framework used to derive the
Witten-Veneziano formula~\cite{wvformula}, which relates the topological
susceptibility $\chi_t$ to the mass of the $\eta{\prime}$.
This formula does work even when the large $N$ limit of the susceptibility is
replaced by its SU(3) value. This fact possibly implies that SU(3) is close
to SU($\infty$).
In Fig.~\ref{fig:chi}
we report results for ${\chi}_{t}^{1/4}/\sqrt{\sigma}$, which has
discretisation artifacts proportional to $a^2 \sigma$. Our fit gives
${\chi}_{t}^{1/4}/\sqrt{\sigma} =0.3739(59) + {0.439}/{N^2}$ for $2\le N
\le 5$.

As $N$ increases and for large values of the lattice size large correlations
appear in the time series of the topological charge.
This phenomenon can be explained by the suppression of instantons of small
size at large $N$, whose distribution as a function of $N$ and of the size
$\rho$ is expected to behave as  $\rho^{(11N/3)  - 5}$~\cite{glue01}.
Our data confirm this behaviour. 
\section{$k$-strings}
\label{seckstrings}
We have studied the ratio $\sigma_k/\sigma_f$ in both D=3+1 and D=2+1
\cite{usstring,usstring2}.
Our results for its zero temperature value at various $N$
and $k$ are reported in Table 1. Those values should be compared
to predictions based on various scenarios for the confining vacuum.
Our data immediately exclude
unbound strings ($\sigma_k \propto k \sigma_f$); also the bag
model \cite{bag} ($\sigma_k \propto \sqrt{[k(N - K)(N +1)]/2N}$) is excluded.
Our results in D=3+1 are in agreement at the $2 \sigma$ level with both the
Casimir scaling~\cite{oldcs} ($\sigma_k/\sigma_f = [k(N-k)]/(N-1)$) and the
MQCD~\cite{mqcd} ($\sigma_k/\sigma_f = \sin(k\pi/N)/\sin(\pi/N))$ predictions.
Other calculations for D=3+1 SU(6) find a better agreement with
MQCD~\cite{Pisa}. Our results in D=2+1 seem to favour Casimir scaling.

At high temperature the spatial $k$-string tensions are found to be closer to
Casimir scaling than to MQCD (the evidence being more striking in D=2+1).~Our
D=3+1 high $T$ results for the spatial $k$-string tensions
agree with a recent model calculation \cite{Chris}.
Other evidences for Casimir scaling in D=2+1  are discussed
in~\cite{usstring2}.
\begin{table}[t]
\label{table:1}
\caption{$\sigma_k/\sigma_f$ in SU($N$) in D dimensions at zero temperature,
for the values of D, $N$ and $k$ shown.} 
\renewcommand{\arraystretch}{1.2} 
\begin{tabular}{|c|c|c|c|}\hline
 D  & $N$  & $k$  & $\sigma_k/\sigma_f$  \\ \hline
 3+1  &  4  & 2  &  1.357(29)  \\
 3+1  &  5  & 2  &  1.583(74)  \\
 2+1  &  4  & 2  &  1.3548(64) \\
 2+1  &  6  & 2  &  1.6160(86) \\
 2+1  &  6  & 3  &  1.808(25)  \\
\hline
\end{tabular}
\end{table}
\section{Conclusions}
\label{secconclusions}
We have reported results from the first modern lattice calculation aimed at
investigating the large $N$ limit of SU($N$) gauge theories in D=3+1.
We have shown that this limit seems to exist. Moreover, the values
of observables like the masses of the lowest-lying glueballs and the
topological susceptibility at finite $N$ are related to their
$N=\infty$ counterpart by a simple $1/N^2$ correction.

We have also investigated the ratios $\sigma_k/\sigma_f$ for
stable $k$-strings. While those ratios seems to fulfill Casimir scaling
in D=2+1 in various regimes and possibly in D=3+1 for the high $T$ spatial
string tensions, our data can not resolve between that prediction and the one
based on MQCD at zero temperature in D=3+1.

\noindent
{\bf Acknowledgements} We are indebted with many colleagues for enlightening
discussions. Correspondence with E. Vicari is gratefully acknowledged.
Numerical simulations have been performed on Compaq Alpha servers partially
funded by PPARC. BL is funded by PPARC under Grant PPA/G/0/1998/00567.
\end{document}